\documentclass[11pt]{article}
\usepackage[margin=1in]{geometry}

\usepackage{fixmath}
\usepackage{bm}
\usepackage{amsbsy}
\usepackage{color}
\usepackage{verbatim}
\usepackage{multirow}
\usepackage{amssymb}
\usepackage{amsthm}
\usepackage{array}
\usepackage{mathtools}
\usepackage{amsmath}
\usepackage{graphicx}
\usepackage{tikz}
\usetikzlibrary{positioning}
\usepackage{xcolor}

\usepackage{thm-restate}

\usepackage[T1]{fontenc}
\usepackage[utf8]{inputenc}
\usepackage{booktabs}
\usepackage[colorlinks=true,linkcolor=blue,citecolor=blue,urlcolor=blue]{hyperref}

\newtheorem{theorem}{Theorem}
\newtheorem{lemma}[theorem]{Lemma}

\newtheorem{observation}[theorem]{Observation}

\newtheorem{reduction}{Reduction Rule}

\newtheorem{branch}{Rule}

\newtheorem{weightedrule}{Rule}

\newcommand{\cost}{c}

\newcommand{\defproblem}[3]{%
  \begin{center}\fbox{\parbox{0.92\textwidth}{%
  #1\\[2pt]\textbf{Input:} #2\\[2pt]\textbf{Output:} #3}}
  \end{center}}

\let\epsilon=\varepsilon

 \usepackage{lineno}

\title
{
Solving (Vertex-Weighted) Directed Multiway Cut Faster Than $2^n$
}

\author{Mingyu Xiao\\[3pt]
\small School of Computer Science and Engineering\\
\small University of Electronic Science and Technology of China\\
\small Chengdu, China\\
\small\texttt{myxiao@uestc.edu.cn}}

\date{}

\begin{document}


\maketitle

\begin{abstract}
In the \textsc{Directed Multiway Cut} problem, we are given a directed graph $G=(V,E)$ and a subset $T\subseteq V$, called the terminal set. The aim is to find a minimum-cardinality set $S\subseteq V\setminus T$, such that after deleting $S$, no directed path exists from one terminal to another terminal in the remaining graph. It had been an open question whether \textsc{Directed Multiway Cut} can be solved faster than the trivial running-time bound $O^*(2^{|V|})$. In this paper, we provide a positive answer to this question, presenting an algorithm with a running-time bound $O(1.9967^{|V|})$. The algorithm can also be modified to solve the weighted version (each vertex having a positive weight) in $O^*(1.9985^{|V|})$ time.

\end{abstract}

\section{Introduction}
The renowned min-cut max-flow theorem establishes that the size of the maximum flow equates to the size of the minimum cut between two vertices in a graph. Remarkably, both the maximum flow and the minimum cut can be computed in polynomial time, showcasing a compelling duality in combinatorics and algorithms.
However, this elegant symmetry fails when we extend to a more generalized problem -- one that involves separating more than two vertices from each other.
The corresponding cut problem, referred to as the \textsc{Multiway Cut} problem~\cite{mwc,Xiao10}, presents significant computational challenges.
In the context of undirected graphs, \textsc{Multiway Cut} becomes NP-complete when the number of terminals in $T$ is three~\cite{mwc,d-mwc}. For directed graphs, the problem turns even more intricate, becoming NP-complete with only two terminals~\cite{d-mwc}.
This paper will delve into the exploration of exact exponential algorithms for the \textsc{Directed Multiway Cut} problem. We will consider both vertex-weighted and unweighted versions.
The weighted version is formally defined as follows.

\defproblem{\textsc{Vertex-Weighted Directed Multiway Cut} (WDMC) }
{Directed graph $G = (V, E)$, a set $T=\{t_1, t_2,\dots, t_l\}\subseteq V$ of $l$ terminal vertices, positive costs $\cost: V\setminus T\to\mathbb{Q}_{>0}$, and a nonnegative budget $k$.}
{Can we delete a set $S\subseteq V\setminus T$ with $\cost(S)=\sum_{v\in S}\cost(v)\leq k$ such that in the remaining graph $G-S$, there is no directed path between any two distinct terminals in $T$?}

For the special case where $\cost (v) = 1$ holds for every nonterminal vertex, the problem becomes the unweighted version and it is called \textsc{Directed Multiway Cut} (DMC). In \textsc{Directed Multiway Cut}, the budget $k$ is the number of vertices allowed to be deleted.  Throughout the paper, $n$ denotes the number of vertices of the input graph, rather than the sum of the vertex costs. Enumerating all subsets of $V\setminus T$ gives a trivial $O^*(2^{n-|T|})$-time\footnote{We use the $O$-notation to hide the constant factor and the $O^*$-notation to hide the polynomial factor.} algorithm for both versions.

In the case of undirected graphs, the unweighted problem is denoted as \textsc{Undirected Multiway Cut}. It is worth noting that \textsc{Multiway Cut} is alternatively referred to as \textsc{Multiterminal Cut} in various literature sources~\cite{mwc,Xiao10}.
Another extension of the minimum cut problem is known as \textsc{Multicut}~\cite{multicut1,multicutESA}. This problem, given a set of terminal pairs, is to disconnect all these terminal pairs by deleting the fewest number of vertices. Depending on whether the graph is directed or undirected, we encounter two distinct problems: \textsc{Directed Multicut} and \textsc{Undirected Multicut}.
\textsc{Multiway Cut} is a specific instance of \textsc{Multicut}. In the former, each pair of terminals will constitute one terminal pair (or two terminal pairs for directed graphs) in the latter.
Both \textsc{Multiway Cut} and \textsc{Multicut} represent significant generalizations of the minimum cut problem, drawing certain attention in the field of graph algorithms.

In the realm of exact algorithms, our goal is not to devise polynomial algorithms for NP-complete problems. Instead, we are interested in determining the fastest possible algorithm within the context of exponential algorithms. For a myriad of NP-complete problems, there are straightforward $O^*(2^n)$-time brute-force algorithms, which enumerate all vertex subsets for graph problems or all assignments for SAT problems, where $n$ is the number of vertices in the graph or the number of variables in SAT. However, despite decades of research, no breakthrough has been made in surpassing this basic running-time bound for several fundamental NP-complete problems. No $O(1.9999^n)$-time algorithm is currently known for the first NP-complete problem, SAT. For general weighted instances, the well-known TSP has not seen any improvement in the exponential base beyond the $O^*(2^n)$-time dynamic programming method developed in the 1960s. \textsc{Multiway Cut} has also been extensively investigated in exact and parameterized algorithms. For \textsc{Undirected Multiway Cut}, Cygan et al. \cite{CyganPPW13} introduced an $O^*(2^k)$-time algorithm, an improvement over the previous result of $O^*(4^k)$~\cite{ChenLL09}.
Using the general framework of designing exact algorithms via parameterized algorithms in~\cite{DBLP:journals/jacm/FominGLS19}, we can directly solve \textsc{Undirected Multiway Cut} in $O^*(1.5^n)$ time. Chitnis et al. \cite{terminalP} proposed an $O(1.4767^n)$-time\footnote{Here we use $O(1.4767^n)$ instead of $O^*(1.4767^n)$ because the actual bound is $O^*(a^n)$ with $a=1.4766\cdots<1.4767$. Thus, $O^*(a^n)\subseteq O(1.4767^n)$.} algorithm for \textsc{Undirected Multiway Cut}.
In the case of \textsc{Undirected Multicut}, Lokshtanov, Saurabh, and Such{\'{y}} \cite{multicutESA} broke the barrier $2^n$ for the first time by providing an $O(1.987^n)$-time algorithm.
When the graph is a directed one, both \textsc{Directed Multiway Cut} and \textsc{Directed Multicut} become considerably more challenging. Before the preliminary version of this paper~\cite{xiaoESA}, whether they could be solved faster than $O(2^n)$ was an open question in the field of exact algorithms~\cite{terminalP,multicutESA}.
Recently, Chu et al.~\cite{ChuEtAl2026} obtained an $O^*(k^{O(k)})=O^*(2^{O(k\log k)})$-time parameterized algorithm for the unweighted \textsc{Directed Multiway Cut} problem, where $k$ is the number of deleted vertices.
However, this bound is not single-exponential in $k$. Combining it with the general framework of designing exact algorithms via parameterized algorithms in~\cite{DBLP:journals/jacm/FominGLS19} does not directly give an $O^*((2-\epsilon)^n)$-time algorithm for a fixed $\epsilon>0$.
In this paper, we show that \textsc{Directed Multiway Cut} can be solved in $O(1.9967^n)$ time, which can also be modified to solve \textsc{Vertex-Weighted Directed Multiway Cut}
in $O^*(1.9985^{n})$ time. An initial version of this paper was presented at ESA 2024~\cite{xiaoESA}, where we only gave the result for unweighted \textsc{Directed Multiway Cut}. In this version, we further simplify the algorithm and also give the result for \textsc{Vertex-Weighted Directed Multiway Cut}.

\textsc{Directed Multiway Cut} admits a 2-factor approximation in polynomial time via the standard vertex-splitting reduction to the edge-deletion version~\cite{DBLP:journals/siamcomp/NaorZ01}. By using the Approximate Monotone Local Search framework described by Esmer et al.~\cite{DBLP:conf/soda/EsmerKMNS24}, we can get an algorithm for  \textsc{Directed Multiway Cut} with the running time better than $2^n$ and the approximation ratio arbitrarily close to 1 (but not 1). The approximation factor can be made arbitrarily close to 1 at the expense of increasing the running time but still keeping it in the form of $O^*((2-\epsilon)^n)$ for some $\epsilon>0$ whenever the approximation factor is fixed and strictly greater than 1.

We should also note the existence of unrestricted versions of \textsc{Multiway Cut} and \textsc{Multicut} where terminals are permitted to be included in the solution set $S$. For these unweighted unrestricted versions, a minimum solution has size at most the number $l$ of terminals, since deleting all terminals is feasible.
This distinction is crucial. When the number $l$ of terminals is small, a simple algorithm that enumerates all vertex subsets of size at most $l$ may already be efficient. Conversely, when the number $l$ of terminals is large, we can first enumerate the vertex subsets of $V\setminus T$ and simplify the problem.
By striking a balance between $|T|$ and $|V|-|T|$, Chitnis et al. \cite{terminalP} demonstrated that the unrestricted version of \textsc{Undirected Multiway Cut} can be solved in $O(1.4767^n)$ time, an improvement over the previous result of $O(1.8638^n)$ in~\cite{DBLP:journals/algorithmica/FominHKPV14}. Furthermore, the unrestricted version of \textsc{Directed Multiway Cut} can be solved in $O(1.6181^n)$ time \cite{terminalP}.

\section{Preliminaries}

Let $G = (V, E)$ denote a directed graph with a set $V$ of vertices and a set $E$ of edges. An edge in $E$, also called an \emph{arc}, is an ordered pair of vertices in $V$. An edge from $a$ to $b$ will be written as $ab$ or $a\to b$. We always use $T$ to denote the set of terminals.
Let $n=|V|$ and $n_0=|V\setminus T|$.
We allow two edges with different directions between two vertices in the graph.

For a vertex $v \in V$, the set of \emph{out-neighbors} of $v$ is defined as $N^{+}(v) := \{ u \in V: vu \in E \}$, and the set of \emph{in-neighbors} of $v$ is defined as $N^{-}(v) := \{ u \in V: uv \in E \}$.
For a vertex subset $V'\subseteq V$, let $N^{+}(V')=\bigcup_{v\in V'} N^{+}(v)\setminus V'$, $N^{-}(V')=\bigcup_{v\in V'} N^{-}(v)\setminus V'$, and $N(V')=N^{+}(V')\bigcup N^{-}(V')$.
For a vertex subset $U$, we also let $N^{+}_U(V')= N^{+}(V')\bigcap U$, $N^{-}_U(V')= N^{-}(V')\bigcap U$, and $N_U(V')= N(V')\bigcap U$.
The subgraph of $G$ induced by a vertex subset $V' \subseteq V$ is denoted by $G[V']$, and we also use $G-V'$ to denote the subgraph $G[V\setminus V']$.
A \emph{path} of length $\ell$ is a sequence of pairwise distinct vertices $(x_{1}, x_{2}, \ldots, x_{\ell+1})$ such that $x_{i}x_{i+1} \in E$ for each $1\leq i \leq \ell$.
If there exists a path from vertex $u$ to vertex $v$, then we say vertex $u$ can \emph{reach} vertex $v$, and conversely, vertex $v$ is \emph{reachable from} vertex $u$.


In our problem, the set of terminals will be denoted by $T=\{t_1,t_2,\dots, t_l\}$.
A path is called a \emph{$T$-path} if the first and last vertex are two distinct terminals.
A set of nonterminal vertices is called a \emph{multiway cut} if there is no $T$-path in the remaining graph after deleting them.
A multiway cut is called minimum if it has minimum cost among all multiway cuts.
In the unweighted problem, the target is to check whether there is a multiway cut of size at most $k$; in the weighted problem, the budget $k$ bounds its total cost.
In our algorithms, we may require a nonterminal vertex $v$ not to be included in the solution. For this case, we will \emph{bypass} the vertex $v$. The definition of this operation is below.
\emph{Bypassing a nonterminal vertex $v$} means adding $a\to b$ for every $a\in N^-(v)$ and $b\in N^+(v)$ with $a\ne b$, and removing $v$. Duplicate arcs are merged, so every bypass leaves the graph without self-loops or duplicate arcs.
We will be utilizing the following simple properties in our algorithms.
\begin{observation}\label{simple1}
Let $v$ lie on no $T$-path in $G$, and put $G'=G-v$ and
$T'=T\setminus\{v\}$. For every $S\subseteq V(G')\setminus T'$, the
set $S$ is a multiway cut for $(G,T)$ if and only if it is one for
$(G',T')$.
\end{observation}
\begin{observation}\label{simple2}
Let $G$ be a directed graph and $G'$ be the graph after reversing the direction of all edges in $G$.
 A vertex set $S$ is a multiway cut in $G$ if and only if $S$ is a multiway cut in $G'$.
\end{observation}
\begin{observation}\label{simple3}
Let $v\in V\setminus T$ and let $G'$ be obtained by bypassing $v$.
For every $S\subseteq V(G')\setminus T$, the set $S$ is a multiway
cut in $G'$ if and only if it is a multiway cut in $G$ that excludes $v$. In particular, if a minimum cut of $G$ excludes
$v$, it is also a minimum cut of $G'$.
\end{observation}

A path through $v$ can be shortened using a new arc. Conversely,
every new arc can be expanded through $v$ into a directed walk, from
which a path can be extracted. This proves the reachability and cost
preservation used in Observation~\ref{simple3}.

We will utilize Observation~\ref{simple1} to simplify the graph by removing certain vertices not involved in any $T$-path. Note that our algorithm may not delete all such vertices.
In one step of our algorithm, we will reverse the direction of all edges, and Observation~\ref{simple2} guarantees the correctness of this step.
In the algorithm, we will also require some nonterminals not in the solution. For this case, we immediately bypass them. Observation~\ref{simple3} guarantees the correctness of this operation.

We also mention that in our algorithms, we will use a measure-and-conquer method to analyze the running time. To do so, we need to introduce a measure for each algorithm.
The unweighted
algorithm uses the measure $w$ and the weighted algorithm uses the measure $\mu$. They are distinct from the cost $\cost(S)$ of a multiway cut $S$.

\section{The Algorithm for the Unweighted Version}\label{sec:unweighted}
In this section, we first design the algorithm for the unweighted problem \textsc{Directed Multiway Cut}. We assume that $\cost(v)=1$ for every nonterminal $v$.
Most of the steps in our algorithm are simple branching rules. To verify the correctness of a branching rule, we only need to demonstrate that solutions exist if and only if at least one can be found in some subbranch.
We will first introduce the main concept and the framework of the algorithm. Then, we will detail the specific steps involved in the algorithm. While introducing each branching rule, we also discuss its correctness and analyze its complexity.

\subsection{Preprocessing}
Initially we discard loops and duplicate arcs.
Our algorithm is a recursive algorithm with several branching operations.
At the start of every recursive call, and after a removal
or reversal, apply the following reductions exhaustively until none
applies. Reject a negative budget $k$ before accepting an empty graph.
\begin{reduction}\label{r-1}
Remove every vertex unreachable from all terminals or
unable to reach any terminal. Remove any weakly connected
component containing at most one terminal.
\end{reduction}
\begin{reduction}\label{r-2}
If an edge joins distinct terminals, reject the instance.
\end{reduction}
\begin{reduction}\label{r-3}
If distinct terminals satisfy $t_i\to v\to t_j$ for a nonterminal vertex $v$,
delete $v$ and decrease $k$ by 1.
\end{reduction}

The correctness of all of the above reduction rules is easy to verify.

\subsection{The Framework}
Now we introduce the framework of the algorithm. In our algorithm, we will always keep an \emph{active} terminal $t$ in the current graph.
Initially, we select an arbitrary terminal as active. During the algorithm, we may also delete an active terminal (when it is not on any $T$-path). For this case, we will select a new terminal as the active terminal.  The active terminal $t$ determines a partition $(A, B, R)$ of the vertex set of the graph, where
\[
 A= \{t\}, \qquad B=N^+(t),\qquad \text{and} \qquad R=V\setminus(A\cup B).
\]
A vertex in $B$ is also called \emph{boundary}. For a boundary vertex $v\in B$, define
\[
U(v)=N^+(v)\cap(R\setminus T), \qquad d(v)=|U(v)|, \qquad \text{and} \qquad U(X)=\bigcup_{v\in X}U(v) ~\text{for}~ X\subseteq B.
\]

In most steps, the algorithm will select one boundary vertex $v\in B$ and branch on $v$ by either excluding it from the solution set $S$ or
including it in the solution set $S$. In the former branch, we bypass $v$, and then $U(v)$ will be moved from $R$ to $B$ (after recomputing the neighborhood of the active terminal $t$).
In the latter branch, we delete $v$ from the graph and decrease the budget $k$ by 1.
The vertices in $R$ may be iteratively moved to $B$ or disconnected from $A\cup B$.
When $N^{+}(A\cup B)=\emptyset$, we may not be able to move vertices in $R$ to $B$ anymore by the above branching operation.
For this case, we will consider two subcases. If it also holds that $N^{-}(A\cup B)= \emptyset$, we delete $A\cup B$ from the graph since
there is only one terminal $t$ in $A\cup B$ (after the preprocessing) and no vertex in $A\cup B$ is in a $T$-path. Next, we will select an arbitrary terminal in the remaining graph as a new active terminal.
If $N^{-}(A\cup B)$ is not an empty set, we will \emph{reverse the direction of all edges} in the graph, keep the same active terminal $t$ and update $B=N^{+}(t)$ and continue our algorithm.

Reversing the direction of all edges is a critical step in our algorithm.
To analyze the algorithm, before performing the reversal, we \emph{mark} all boundary vertices in $B$ and use $Y$ to denote the set of marked vertices. In the algorithm analysis, we will distinguish the complexity between marked and unmarked vertices. When there exist marked vertices in the graph (it will be proved later that when they exist there always exist marked vertices in $B$), our algorithm will preferentially choose marked vertices in $B$ for branching.

\paragraph{Algorithm 1.} (Recursive framework)

\emph{Preprocess.} Apply Reduction Rules~\ref{r-1}, \ref{r-2},
and~\ref{r-3}. Reject a negative budget and accept an empty graph. Restrict $T$
and $Y$ to surviving vertices that are not deleted or bypassed. If no active terminal is selected or it has
been removed, select any surviving terminal as the new active terminal;
otherwise retain the current active terminal.

\emph{Select a branching rule.} If there are no marked vertices ($Y=\emptyset$), use the first applicable rule
among Rules~\ref{b-1}--\ref{b-3}. If none applies, remove $A\cup B$ from the graph when $N^-(A\cup B)=\emptyset$,
and otherwise apply Rule~\ref{b-4} to reverse the direction of edges. In either case return to preprocessing.

If there are some marked vertices ($Y\ne\emptyset$), use the first applicable rule among
Rules~\ref{b-5}--\ref{b-7}. Lemma~\ref{marked} guarantees that some
marked vertex belongs to $B$ and at least one of these rules can be applied.

\emph{Recurse.} Recursively solve the
resulting instance.
Accept if some branch accepts, and reject if all
branches reject.
Each time when the graph changes, we will update the set $A$, $B$, $R$, $U(B)$ and so on. For the convenience of algorithm analysis, we may color the vertices in the graph, and we will also update the coloring each time the graph changes.

The weighted algorithm in Section~\ref{sec:weighted} uses the same
framework, replacing Rules~\ref{b-5} and~\ref{b-6} by Rules~\ref{wr-1} and~\ref{wr-2}, respectively, and retaining Rule~\ref{b-7}. Each paid deletion subtracts the vertex's input cost rather than one from the budget.

\subsection{Main Steps of the Algorithm}\label{sec:unmarked-rules}

We will use a measure-and-conquer method to analyze our algorithm.
The measure-and-conquer method was introduced by Fomin et al.~\cite{DBLP:journals/jacm/FominGK09}. It is a powerful tool to analyze exact algorithms.
In the measure-and-conquer method, we will set \emph{analysis weights} to different types of vertices to distinguish their contribution to the complexity. We will make sure that the weight of each vertex is at most 1. Then the total analysis weight of all vertices is not greater than the number of vertices in the graph. Thus, we can use the sum of analysis weights as the measure to analyze the algorithm and get a running time bound with respect to the number of vertices in the graph.
We will analyze each branching operation of the algorithm after introducing it.

\paragraph{Analysis measure.}
To elucidate how we assign analysis weights to the vertices, we use two different labels: coloring and marking. The vertices will be colored in either red or blue by using the following \emph{color principle}, meaning a vertex can be either one of these colors or uncolored. At a certain step in our algorithm as mentioned above, we will mark some vertices, making them either marked or unmarked.

\emph{Color principle.}
Boundary vertices $v\in B$ are colored red if $U(v)=\emptyset$ and blue otherwise; vertices in $R$ are
uncolored. Note that neighborhoods and colors are recomputed when the graph changes.

\emph{Weight setting.} The following weight setting will be used in our analysis. Each vertex will receive an analysis weight and we have five different analysis weights for vertices: $1, \alpha=0.958234, \beta=0.943555, \gamma=0.901789,$ and 0.
Terminals always have analysis weight 0. Nonterminal vertices are assigned an analysis weight according to their labels, as shown in Table~\ref{table-weight}.

\begin{table}[htbp]
\centering
\begin{tabular}{lccc}
\toprule
& blue in $B$ & red in $B$ & vertex in $R$\\
\midrule
unmarked & $\alpha$ & $\beta$ & $1$\\
marked & $\gamma$ & $\gamma$ & $\beta$\\
\bottomrule
\end{tabular}
\caption{The analysis weight for nonterminals, where $\alpha=0.958234$, $\beta=0.943555$, and $\gamma=0.901789$.}
\label{table-weight}
\end{table}

Let $w$ be the sum of the analysis weights and we will use it as the measure to evaluate the running time. Initially $w\le n_0$, since terminals have an analysis weight 0 and nonterminals have an analysis weight $\leq 1$.
For a recurrence with $r$ branches in which the measure $w$ decreases by $a_1,\ldots,a_r$, respectively, a bound $b^w$
on the number of leaves follows from $\sum_j b^{-a_j}\le1$, provided
all nonbranching operations are measure-nonincreasing. Polynomial work
per state is accounted for separately. The detailed background about how to analyze recurrences can be found in~\cite{DBLP:series/txtcs/FominK10}.

\paragraph{Branch and reversal rules.}
The next four rules work for the case where there is no marked vertex ($Y=\emptyset$).

\begin{branch}\label{b-1}
If $Y=\emptyset$ and there is a vertex $v\in B$ with $d(v)\ge3$,
branch on $v$ by either deleting it or bypassing it.
The budget $k$ decreases by $1$ and $0$, respectively.
\end{branch}

The correctness of Rule~\ref{b-1} follows by considering whether $v$ belongs to the solution cut $S$. In the first branch, the blue vertex $v$ is deleted from the graph, and the measure $w$ decreases by at least $\alpha$.
In the second branch, $v$ is bypassed. This creates
$t\to u$ for every $u\in U(v)$, so all these vertices enter $B$.
Thus $d(v)$ uncolored vertices become red or blue. Since $\alpha > \beta$, the measure $w$ decreases in total by at least
$\alpha + d(v) (1-\alpha)$, where $d(v)\geq3$. We use $C(w)$ to denote the maximum number of leaves in the search tree for a state of total analysis weight at most $w$.
This yields the recurrence
\begin{equation}
C(w)\leq C(w-\alpha)+ C(w-(3-2\alpha)),
        \end{equation}
whose branching factor is below 1.9736.

\begin{branch}\label{b-2}
If $Y=\emptyset$ and there is a vertex $v\in B$ with $d(v)=1$, write
$U(v)=\{u\}$. Branch on $u$ by either bypassing it or
deleting it. The budget $k$ decreases by $0$ and $1$, respectively.
\end{branch}

The correctness of Rule~\ref{b-2} follows by considering whether $u$ belongs to the solution cut $S$. In the first branch, the uncolored vertex $u$ is bypassed, and the measure $w$ decreases by 1.
In the second branch, the uncolored vertex $u$ is deleted from the graph, leaving the vertex $v$ with $U(v)=\emptyset$. Its color therefore changes from blue to red, and its weight changes from $\alpha$ to $\beta$. The measure $w$ decreases by at least $1+(\alpha-\beta)$.
This yields the recurrence
\begin{equation}
C(w)\leq  C(w-1) + C(w-(1+\alpha-\beta)),
        \end{equation}
whose branching factor is below 1.9900.

\begin{branch}\label{b-3}
If $Y=\emptyset$ and there is a vertex $v\in B$ with $d(v)=2$, write
$U(v)=\{u_1,u_2\}$. Use the following three branches:
\begin{enumerate}
\item Bypass $u_1$.
\item Delete $u_1$, and bypass $u_2$.
\item Delete both $u_1$ and $u_2$.
\end{enumerate}
The budget decreases by $0$, $1$, and $2$, respectively.
\end{branch}

The correctness of Rule~\ref{b-3} follows by considering whether $u_1$ belongs to the solution cut $S$.
If $u_1$ is deleted, we further branch on whether $u_2$ is deleted. This yields three branches.
In the first branch, $u_1$ is excluded from the solution. The uncolored vertex $u_1$ is bypassed, and the measure $w$ decreases by at least 1. In the second branch, the uncolored vertex $u_1$ is deleted, the uncolored vertex $u_2$ is bypassed, and the measure $w$
decreases by at least 2. In the third branch, the two uncolored vertices $u_1$ and $u_2$ are deleted, the vertex $v$ changes from blue to red, its weight changes from $\alpha$ to $\beta$, and the measure $w$ decreases by at least $2+(\alpha-\beta)$.
This yields the recurrence
\begin{equation}\label{bo1}
C(w)\leq C(w-1)+C(w-2)+C(w-(2+\alpha-\beta)),
        \end{equation}
whose branching factor is at most 1.99664.

\paragraph{Some properties after Rule~\ref{b-3}.}
If none of Rules~\ref{b-1}--\ref{b-3} applies, every vertex of $B$ is red.
There is no arc from $B$ to a vertex in $R$, since $U(B)=\emptyset$.
There is no arc from $B$ to another terminal, by Reduction Rule~\ref{r-3}.
Consequently, $A\cup B$ has no outgoing arc to its complement.
Every vertex of $B$ can reach a terminal after preprocessing, and the
only terminal in $A\cup B$ is $t$. Hence every vertex of $B$ reaches $t$, and $G[A\cup B]$ is strongly connected. If $A\cup B$ also has no incoming arc,
it will be removed by preprocessing. Otherwise Rule~\ref{b-4} marks the current
$B$ and reverses all arcs, keeping the current active terminal $t$.

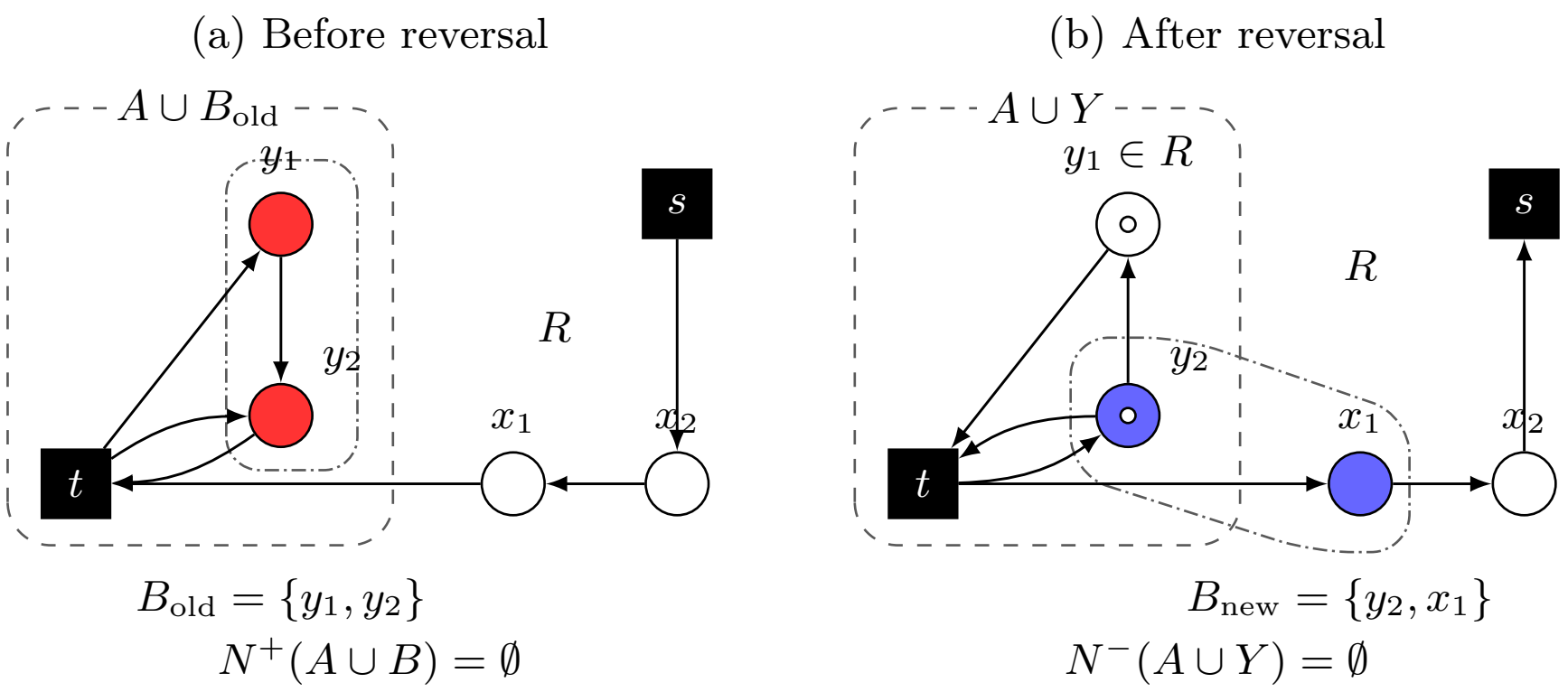
\begin{figure}[htbp]
\centering
\begingroup
\newcommand{\NormalizedReversePanel}[1]{%
  \tikzset{vertex/.style={circle,draw=black,line width=.5pt,
      fill=white,minimum size=4.6mm,inner sep=0pt},
    terminal/.style={rectangle,draw=black,line width=.5pt,
      fill=black,text=white,minimum size=5mm,inner sep=0pt},
    closedregion/.style={draw=black!65,dashed,line width=.5pt},
    boundaryregion/.style={draw=black!65,dash pattern=on 3pt off 1pt on .6pt off 1pt,
      line width=.5pt},
    arc/.style={->,line width=.55pt}}
  \draw[closedregion,rounded corners=9pt] (.10,.20) rectangle (2.92,3.40);
  \ifnum#1=0\relax
    \draw[boundaryregion,rounded corners=7pt] (1.70,.75) rectangle (2.66,3.02);
    \node[fill=white,inner sep=2pt] at (1.50,3.40) {$A\cup B_{\mathrm{old}}$};
    \node at (2.10,-.20) {$B_{\mathrm{old}}=\{y_1,y_2\}$};
    \node at (4.10,1.80) {$R$};
    \node[vertex,fill=red!80,label=above:{$y_1$}] (nyone) at (2.10,2.55) {};
    \node[vertex,fill=red!80,label=above right:{$y_2$}] (nytwo) at (2.10,1.15) {};
    \node[vertex,label=above:{$x_1$}] (nxone) at (3.80,.65) {};
  \else
    \draw[boundaryregion,rounded corners=8pt]
      (1.68,.72) -- (3.46,.15) -- (4.16,.15) -- (4.16,1.15)
      -- (2.50,1.72) -- (1.68,1.72) -- cycle;
    \node[fill=white,inner sep=2pt] at (1.50,3.40) {$A\cup Y$};
    \node at (3.65,-.20) {$B_{\mathrm{new}}=\{y_2,x_1\}$};
    \node at (3.80,2.25) {$R$};
    \node[vertex,label=above:{$y_1\in R$}] (nyone) at (2.10,2.55) {};
    \node[vertex,fill=blue!60,label=above right:{$y_2$}] (nytwo) at (2.10,1.15) {};
    \node[vertex,fill=blue!60,label=above:{$x_1$}] (nxone) at (3.80,.65) {};
  \fi
  \node[terminal] (nt) at (.60,.65) {$t$};
  \node[terminal] (ns) at (5.00,2.70) {$s$};
  \node[vertex,label=above:{$x_2$}] (nxtwo) at (5.00,.65) {};
  \foreach \source/\target in {nt/nyone,nyone/nytwo,ns/nxtwo,nxtwo/nxone,nxone/nt}
    {\ifnum#1=0\relax
      \draw[arc] (\source) -- (\target);
    \else
      \draw[arc] (\target) -- (\source);
    \fi}
  \foreach \source/\target in {nt/nytwo,nytwo/nt}
    {\ifnum#1=0\relax
      \draw[arc] (\source) to[bend left=17] (\target);
    \else
      \draw[arc] (\target) to[bend right=17] (\source);
    \fi}
  \ifnum#1=1\relax
    \foreach \markedvertex in {nyone,nytwo}
      \draw[line width=.5pt,fill=white] (\markedvertex) circle (.055);
  \fi
}
\begin{tikzpicture}[x=1cm,y=1cm,>=latex,font=\footnotesize]
  \begin{scope}
    \NormalizedReversePanel{0}
    \node at (2.75,3.92) {(a) Before reversal};
    \node at (2.75,-.65) {$N^+(A\cup B)=\emptyset$};
  \end{scope}
  \begin{scope}[xshift=6.20cm]
    \NormalizedReversePanel{1}
    \node at (2.75,3.92) {(b) After reversal};
    \node at (2.75,-.65) {$N^-(A\cup Y)=\emptyset$};
  \end{scope}
\end{tikzpicture}
\endgroup
\caption{A state before and after Rule~\ref{b-4}, with active
terminal $t$. Squares are terminals and circles are nonterminals. Before reversal, the old set of boundary vertices
$B_{\mathrm{old}}=\{y_1,y_2\}$ is red and $A\cup B_{\mathrm{old}}$
has no outgoing arc to its complement. The step marks
$Y=\{y_1,y_2\}$ (small rings), reverses all arcs so that after the reversal $A\cup Y$ has no incoming arc,
and recomputes the new set of boundary vertices as $B_{\mathrm{new}}=\{y_2,x_1\}$, whose
vertices are blue. Thus the marked vertex $y_1$ is in $R$, the boundary
vertex $x_1$ is unmarked, and $B_{\mathrm{new}}\ne Y$.}
\label{reverse}
\end{figure}

\begin{branch}\label{b-4}
If $Y=\emptyset$, none of Rules~\ref{b-1}--\ref{b-3} applies, and
$N^-(A\cup B)\ne\emptyset$, then mark every vertex in the old $B$ (let $Y=B$), reverse all arcs in the graph, and keep the current active terminal $t$ and update the vertex set $B=N^+(t)$ in the new orientation.
\end{branch}
The set $Y$ consists of the surviving marked vertices. Immediately
after Rule~\ref{b-4}, it equals the old boundary, but the new $B$ need
not equal $Y$. Some marked vertices can lie in $R$, and some new
boundary vertices can be unmarked. See Figure~\ref{reverse} for an illustration.

\paragraph{The marked phase.}\label{sec:marked-rules}
The following several rules will be used only when there are some marked vertices ($Y\neq \emptyset$).
We will call them \emph{marked rules}.
Before giving the marked rules, we prove the reachability invariant
created by Rule~\ref{b-4}. It is the reason that retaining selected
marked vertices can preserve an optimal cut. Once a marked vertex
survives, the procedure uses only the marked rules until all marks have
been removed; it does not reverse the graph during this phase.
The next lemmas also establish measure monotonicity.

\begin{lemma}\label{Y-vertex}
Immediately after Rule~\ref{b-4}, no terminal other than $t$ can reach
a marked vertex in $Y$.
\end{lemma}
\begin{proof}
Before reversal, $A\cup B$ has no outgoing arc. After reversal,
$A\cup Y$ has no incoming arc from outside and contains no
terminal other than $t$.
\end{proof}
Note that after deleting or bypassing a vertex, no terminal other than $t$ can reach
a marked vertex in $Y$.

\begin{lemma}\label{connect-1}
Immediately after Rule~\ref{b-4}, for every $u\in A\cup Y$, there is a
path from $t$ to $u$ and a path from $u$ to $t$ inside $G[A\cup Y]$.
Every path from $t$ to a marked vertex lies wholly in $A\cup Y$.
Moreover, after any subsequent operations of deleting and bypassing a vertex followed by exhaustive
preprocessing, each surviving marked vertex is reachable from $t$
by a path containing only $t$ and surviving marked vertices.
\end{lemma}
\begin{proof}
Before reversal, $G[A\cup B]$ is strongly connected, as proved above.
Reversal preserves strong connectivity, and marking gives the
set $Y=B$ from the previous orientation. The absence of external
incoming arcs, established in Lemma~\ref{Y-vertex}, also shows that a
path starting at $t$ and ending in $Y$ cannot leave this set and reenter
it. This property persists under deletion and bypassing. A surviving
marked vertex cannot be reachable from any other terminal. If it were
also unreachable from $t$, preprocessing would remove it. Its path from
$t$ must therefore lie inside the surviving $A\cup Y$. If the
component of $t$ is removed, every marked vertex becomes
unreachable from all terminals and is removed by preprocessing.
\end{proof}

\begin{lemma}\label{lem-4}
Rule~\ref{b-4} does not increase the measure $w$.
\end{lemma}
\begin{proof}
Before reversal, all boundary vertices are unmarked red vertices of
weight $\beta$. After marking and recomputing the boundary, each has
weight $\beta$ if it is in $R$ and $\gamma$ if it is in $B$.
Every newly exposed unmarked boundary vertex previously had weight
$1$ and now has weight $\alpha$ or $\beta$. Terminals have weight zero.
Since $\gamma<\beta<\alpha<1$, marking, reversal, and recoloring do not
increase $w$.
\end{proof}

\begin{lemma}\label{marked}
If no marked vertex belongs to $B$, then after preprocessing no
marked vertex remains in the graph.
\end{lemma}
\begin{proof}
Suppose there is a marked vertex $v$ not in $B$ in the graph after preprocessing. By Lemmas~\ref{Y-vertex} and
\ref{connect-1}, it is reachable from no terminal other than $t$, and there is a path from
$t$ to $v$ containing only $t$ and surviving marked vertices. The
vertex immediately after $t$ on that path is marked and lies in $N^+(t)=B$, a
contradiction. If $t$ was removed, preprocessing removes all remaining
marked vertices as unreachable from every terminal.
\end{proof}

\begin{lemma}\label{yellow}
For each marked
$v\in B$, there is a minimum-cardinality multiway cut $S^*$ such that
either $v\notin S^*$ or $|U(v)\setminus S^*|\ge2$.
\end{lemma}
\begin{proof}
Let $v\in B$ be marked, put $U=U(v)$, and suppose a minimum cut $S'$
contains $v$ and satisfies $|U\setminus S'|\le1$. Define
$S^*=(S'\cup U)\setminus\{v\}$. If $G-S^*$ contained a $T$-path,
choose a simple such path. It must contain $v$, because $S^*$ restores
only $v$, while all other vertices of $S'$ remain removed. By
Lemma~\ref{Y-vertex}, this path starts at $t$. The vertex immediately
after $v$ cannot be in $U$, since $U$ is deleted, and cannot be $t$,
since the path is simple. It cannot be another terminal either: the
arcs $t\to v$ and $v\to t_j$ would have forced the deletion of $v$ during preprocessing.
It is therefore another vertex $b\in B$. Replace the path prefix
ending at $b$ by the arc $t\to b$. The remaining suffix avoids $v$ and
all vertices of $S'$, so this gives a terminal path in $G-S'$, a
contradiction. Thus $S^*$ is a cut, and
$|S^*|=|S'|-1+|U\setminus S'|\le|S'|$. It is another minimum cut and
does not contain $v$, as required.
\end{proof}

By Lemma~\ref{yellow}, we know that if $|U(v)|\leq 1$, then there is
always a minimum multiway cut not containing $v$.

\begin{branch}\label{b-5}
If there is a marked vertex $v\in B$ with $d(v)\leq1$, bypass
$v$.
\end{branch}

Next, we consider a marked vertex $v\in B$ with $2\leq d(v)\leq4$.
The three low-degree branching rules can be written as a single rule.
Write $U(v)=\{u_1,\ldots,u_d\}$ in an arbitrary order, where $d=d(v)$.

\begin{branch}\label{b-6}
If there is a marked vertex $v\in B$ with $2\leq d(v)\leq4$, use the
following $d$ branches, choosing such a vertex with minimum $d(v)$:
\begin{enumerate}
\item Bypass $v$.
\item For each $i=1,\ldots,d-2$, delete $v,u_1,\ldots,u_{i-1}$ and
bypass $u_i$.
\item Delete $v,u_1,\ldots,u_{d-2}$ and bypass both
$u_{d-1}$ and $u_d$.
\end{enumerate}
The budget $k$ decreases by $0$, $i$, and $d-1$, respectively.
The prefix $u_1,\ldots,u_{i-1}$ is empty when $i=1$. When $d=2$,
there are no branches of the second type, and the final branch deletes
only $v$ and bypasses both vertices of $U(v)$.
\end{branch}

By Lemma~\ref{yellow}, we can choose a minimum multiway cut $S^*$ such
that either $v\notin S^*$ or at least two vertices in $U(v)$ are not in
$S^*$. If $v\notin S^*$, the first branch applies. Otherwise,
$v\in S^*$ and at least two vertices of $U(v)$ are outside $S^*$.
If some vertex among $u_1,\ldots,u_{d-2}$ is not in $S^*$, let
$u_i$ be the first such vertex. Then $v,u_1,\ldots,u_{i-1}$ are in
$S^*$, and the branch indexed by $i$ applies. Otherwise, all of
$u_1,\ldots,u_{d-2}$ belong to $S^*$, so both $u_{d-1}$ and $u_d$
must be outside $S^*$, as specified by the final branch. Thus the rule
preserves a minimum multiway cut.

In the first branch, the marked blue vertex $v$ is
bypassed, and the measure $w$ decreases by at least $\gamma$. Every
vertex of $U(v)$ is an uncolored vertex of weight at least $\beta$;
these vertices may already be marked. In the branch indexed by $i$,
the marked blue vertex $v$ and $i-1$ uncolored vertices are deleted,
and one uncolored vertex is  bypassed. Thus, the measure
$w$ decreases by at least $\gamma+i\beta$. In the final branch, $v$
and $d-2$ uncolored vertices are deleted, and two uncolored vertices
are bypassed. The measure $w$ decreases by at least
$\gamma+d\beta$. We get a recurrence relation
\begin{equation}\label{eq:unit-lowdegree}
C(w)\leq C(w-\gamma)+\sum_{i=1}^{d-2}C(w-(\gamma+i\beta))
       +C(w-(\gamma+d\beta)),\qquad 2\leq d\leq4,
\end{equation}
where the sum in the middle item is empty for $d=2$.
For $d=2$ and $d=3$, the branching factors are below $1.5166$ and
$1.8453$, respectively. For $d=4$, we get
\begin{equation}\label{bo2}
C(w)\leq C(w-\gamma)+C(w-(\gamma+\beta))
       +C(w-(\gamma+2\beta))+C(w-(\gamma+4\beta)),
\end{equation}
whose branching factor is at most $1.99664$.


\begin{branch}\label{b-7}
If there is a marked vertex $v\in B$ with $d(v)\ge5$, branch on $v$
by either bypassing it or deleting it.
The budget $k$ decreases by $0$ and $1$, respectively.
\end{branch}
In the first branch, the marked blue vertex $v$ is bypassed, and the number of uncolored vertices that become red or blue is at least $|U(v)|\geq 5$.
When an unmarked uncolored vertex becomes red or blue, the measure $w$ decreases by at least $\min \{ 1-\alpha, 1-\beta\}=1-\alpha$.
When a marked uncolored vertex becomes red or blue, the measure $w$ decreases by at least $\beta-\gamma$. Set $\Delta=\min \{1-\alpha, \beta-\gamma\}=0.041766$.
The measure $w$ therefore decreases by at least $\gamma+ 5\Delta$.
In the second branch, the marked blue vertex $v$ is deleted, and the measure $w$ decreases by $\gamma$. This yields the recurrence
\begin{equation}\label{bo3}
C(w)\leq C(w-(\gamma+5\Delta))+ C(w-\gamma),
        \end{equation}
whose branching factor is at most 1.99664.

\subsection{The Correctness and Running Time}\label{sec:unit-runtime}

\begin{lemma}\label{lem:correctness}
Algorithm~1 always terminates
and accepts if and only if the input has a multiway cut of size at most $k$.
\end{lemma}
\begin{proof}
Consider a nonempty state after exhaustive preprocessing. A surviving
terminal exists, so the active terminal can be selected. If $Y=\emptyset$,
either some boundary vertex is blue, in which case its degree is $1$,
$2$, or at least $3$ and one of Rules~\ref{b-1}--\ref{b-3} applies,
or every boundary vertex is red. In the latter case, the argument above
Rule~\ref{b-4} justifies removing $A\cup B$ or reversing all arcs.
If $Y\ne\emptyset$, Lemma~\ref{marked} supplies a marked vertex in $B$.
Its degree is at most $1$, between $2$ and $4$, or at least $5$, so one
of Rules~\ref{b-5}--\ref{b-7} applies. Thus every nonempty preprocessed
state has a prescribed next operation.
No operation introduces vertices. Every recursive
child deletes or bypasses at least one vertex, and every removal decreases the number of vertices. A reversal with nonempty old
$B$ creates a nonempty marked set; before another reversal, a vertex
must be removed by preprocessing or a marked rule. A reversal with
empty old $B$ gives a nonempty new $B$, since $N^-(\{t\})\ne\emptyset$.
Without a vertex removal, any next reversal therefore has nonempty old
$B$. Hence at most two reversals occur between consecutive vertex
removals. Preprocessing terminates for the same reason, and the entire
finitely branching search is finite.

It remains to prove correctness. The reductions and the removal of
$A\cup B$ preserve the existence of a cut within the remaining budget,
and Observation~\ref{simple2} justifies reversal. Bypassing a vertex
preserves terminal reachability among the remaining vertices and
therefore represents retaining that vertex. Thus a cut returned by a
child, together with the paid deletions, is a cut of its parent within
the parent's budget. Conversely, Rules~\ref{b-1}--\ref{b-3} and
Rule~\ref{b-7} exhaust the relevant choices, while
Lemma~\ref{yellow} shows that Rules~\ref{b-5} and~\ref{b-6} preserve
a minimum cut in at least one child. Hence a feasible parent has a
feasible child. A negative budget or an arc between distinct terminals
correctly forces rejection, whereas an empty graph needs no further
deletion. Induction over the finite search tree proves both directions
of the claim.
\end{proof}

\paragraph{Measure monotonicity.}
\label{par:measure-overhead}
We also verify that the operations between branchings do not increase
the measure. While $t$ is fixed, a surviving boundary vertex stays in
$B$. Deletion cannot turn a red vertex blue. Bypassing an $R$ vertex
cannot do so either, since a red vertex has no nonterminal out-neighbor
in $R$. When a boundary vertex is bypassed, all its nonterminal
out-neighbors in $R$ enter $B$, so any new nonterminal out-neighbor of a red
boundary vertex is also in $B$. Thus surviving red vertices remain
red, while moving an $R$ vertex into $B$ never increases its analysis
weight. Harmless removals only decrease the measure, and
Lemma~\ref{lem-4} handles reversal. If $t$ is removed, its component,
including its boundary, is removed. After preprocessing reaches a
fixed point, no old mark survives, by Lemma~\ref{Y-vertex} and
Reduction Rule~\ref{r-1}. Every remaining vertex is therefore an
unmarked $R$ vertex before a new active terminal is selected, and this
selection cannot increase the measure.

 Each state is processed in polynomial time, so the
recurrences bound the number of leaves by $1.99664^w$ and the total
running time by $O^*(1.99664^{n_0})$. Processing branches in depth-first
order uses polynomial space. The same monotonicity and overhead
arguments apply to the weighted measure $\mu$, whose analysis weights
have the same ordering.

Since $n_0=n-|T|$, the following bound holds. Under the usual
unweighted encoding, $k$ may be capped at $n_0$. Moreover,
$1.99664<1.9967$, so polynomial overhead in the original vertex count
$n$ can be absorbed by the larger base, giving the $O(1.9967^n)$
bound stated in the introduction.
\begin{theorem}\label{thm:unweighted}
\textsc{Directed Multiway Cut} can be solved deterministically in
$O^*(1.9967^{n-|T|})$ time and polynomial space.
\end{theorem}

For the values of $\alpha, \beta$ and $\gamma$, we decide them by the following way.
We set $\alpha, \beta$ and $\gamma$ as variables and generate a constraint for each of above branching recurrences to get a quasiconvex program.
By solving this quasiconvex program according to the method introduced in~\cite{eQP}, we get a bound of $O(1.99664^w)$  on the number of leaves by setting $\alpha=0.958234$, $\beta=0.943555$ and $\gamma=0.901789$  for our problem.

\section{The Vertex-Weighted Version}\label{sec:weighted}
Now we consider the vertex-weighted version, where each vertex has a positive deletion cost.
The framework and most steps of the algorithm are the same as that of the algorithm for the unweighted version.
Only the marked rules~\ref{b-5} and~\ref{b-6} are changed. This is necessary because
Lemma~\ref{yellow} does not work for the weighted version. Each paid deletion of a vertex $v$ decreases the budget by $\cost(v)$; bypassing and harmless removals do not change the budget. The preprocessing, reversal, and reachability invariants are independent of deletion costs.
In the analysis of the running time bound, the vertex analysis weights will also need to be updated.

\subsection{New Branching Rules}\label{sec:weighted-rules}
The exchange in Lemma~\ref{yellow} does not hold
for arbitrary vertex costs. For the vertex-weighted version, the following weaker statement suffices.
\begin{lemma}\label{lem:exchange}
Let $v\in B$ be
marked. If a feasible cut $S$ contains $v$ and all of $U(v)$, then
$S\setminus\{v\}$ is also feasible. A minimum-cost solution therefore
never contains $\{v\}\cup U(v)$.
\end{lemma}
\begin{proof}
Suppose restoring $v$ creates a simple path between distinct terminals.
It contains $v$ and must start at $t$ by Lemma~\ref{Y-vertex}. The vertex
immediately after $v$ cannot be in $U(v)$, because the vertices of that set remain deleted,
it cannot be $t$, because the path is simple, and it cannot be another
terminal, because preprocessing excluded $t\to v\to t_j$. The next vertex must therefore be
another vertex $b\in B$. Replace the prefix ending at $b$ by $t\to b$.
The suffix avoids $v$, so a terminal path already existed in $G-S$, a
contradiction. Since costs are positive, additionally deleting $v$ strictly increases the cost.
\end{proof}

Assume that $Y\neq \emptyset$. We still have three marked rules. The first two are presented below and the third one is still Rule \ref{b-7}.

\begin{weightedrule}\label{wr-1}
If there is a marked vertex $v\in B$ with $d(v)=0$, bypass $v$.
\end{weightedrule}

\begin{weightedrule}\label{wr-2}
If there is a marked vertex $v\in B$ with $1\leq d(v)\leq4$,
write $U(v)=\{u_1,\ldots,u_d\}$ in an arbitrary order, where $d=d(v)$, and use the following $d+1$ branches:
\begin{enumerate}
\item For each $i=1,\ldots,d$, delete $u_1,\ldots,u_{i-1}$ and bypass $u_i$. The budget $k$ decreases by
$\sum_{j=1}^{i-1}\cost(u_j)$; the prefix is empty when $i=1$.
\item Delete all of $U(v)$ and bypass $v$.
The budget $k$ decreases by $\sum_{j=1}^{d}\cost(u_j)$.
\end{enumerate}
\end{weightedrule}

Rule~\ref{wr-2} enumerates the first retained neighbor (not in the solution cut) in the chosen
order. If all neighbors are deleted, Lemma~\ref{lem:exchange} permits
retaining $v$, which gives the final branch. Rule~\ref{wr-1} is the
same exchange when $U(v)$ is empty.
Thus these rules preserve an optimal
solution in every state.

The termination argument of Lemma~\ref{lem:correctness} is independent
of deletion costs and applies unchanged. Its coverage argument also
applies: if marked vertices remain, Lemma~\ref{marked} supplies a
marked vertex in $B$, whose degree is $0$, between $1$ and $4$, or at
least $5$. Thus Rule~\ref{wr-1}, Rule~\ref{wr-2}, or Rule~\ref{b-7}
applies, respectively. The reductions and unmarked rules remain valid.
Lemma~\ref{lem:exchange} justifies the new marked rules, and
Rule~\ref{b-7} exhausts deletion and retention. Only paid deletions are
charged to the budget, by subtracting their total input cost; bypassing
and harmless removals do not change the budget. Consequently, induction
over the finite search tree proves that the weighted algorithm accepts
if and only if a multiway cut of total cost at most $k$ exists.

\subsection{The Running Time Analysis}
In this section, we analyze the running time. We will set different analysis weights to vertices.

\emph{Weight setting.} Terminals always have analysis weight 0. Let
$\alpha_{\mathrm w}=0.983267$, $\beta_{\mathrm w}=0.976613$, $\gamma_{\mathrm w}=0.959880$, and
$\Delta=\min\{1-\alpha_{\mathrm w},\beta_{\mathrm w}-\gamma_{\mathrm w}\}=0.016733$. Nonterminal vertices are assigned an analysis weight according to their labels, as shown in Table~\ref{table-weight2}.

\begin{table}[htbp]
\begin{center}
\begin{tabular}{lccc}
\toprule
& blue in $B$ & red in $B$ & vertex in $R$\\
\midrule
unmarked & $\alpha_{\mathrm w}$ & $\beta_{\mathrm w}$ & $1$\\
marked & $\gamma_{\mathrm w}$ & $\gamma_{\mathrm w}$ & $\beta_{\mathrm w}$\\
\bottomrule
\end{tabular}
\end{center}
\caption{The analysis weight for nonterminals in the vertex-weighted version, where $\alpha_{\mathrm w}=0.983267$, $\beta_{\mathrm w}=0.976613$, and $\gamma_{\mathrm w}=0.959880$.}
\label{table-weight2}
\end{table}

Let $\mu$ be the sum of the analysis weights of all vertices, initially at most $n_0=n-|T|$.
We will use $\mu$ as the measure to evaluate the running time.
Let $C(\mu)$ denote the maximum number of leaves in the search tree for a weighted state of total analysis weight at most $\mu$.

 The three recurrences for Rules~\ref{b-1}--\ref{b-3} are
\begin{align}
C(\mu)&\le C(\mu-\alpha_{\mathrm w})+C(\mu-(3-2\alpha_{\mathrm w})),\label{eq:u3}\\
C(\mu)&\le C(\mu-1)+C(\mu-(1+\alpha_{\mathrm w}-\beta_{\mathrm w})),\label{eq:u1}\\
C(\mu)&\le C(\mu-1)+C(\mu-2)+C(\mu-(2+\alpha_{\mathrm w}-\beta_{\mathrm w})).\label{eq:u2}
\end{align}

Next, we consider the three marked rules. Rule~\ref{wr-1} removes a vertex without branching and does not increase $\mu$.

We next consider Rule~\ref{wr-2}. Every vertex of $U(v)$ lies in $R$
and has analysis weight at least $\beta_{\mathrm w}$, whether marked or
unmarked. In branch $i$, deleting $i-1$ such vertices and bypassing
$u_i$ decreases $\mu$ by at least $i\beta_{\mathrm w}$. The final branch
deletes all $d$ neighbors and bypasses the marked boundary vertex $v$,
decreasing $\mu$ by at least $d\beta_{\mathrm w}+\gamma_{\mathrm w}$.
Other recolorings do not increase the measure. Hence
\begin{equation}\label{eq:mlow}
C(\mu)\le\sum_{i=1}^{d}C(\mu-i\beta_{\mathrm w})
       +C(\mu-d\beta_{\mathrm w}-\gamma_{\mathrm w}),
\qquad 1\le d\le4.
\end{equation}
The last rule is still Rule \ref{b-7}.
In Rule~\ref{b-7}, the first branch, which bypasses $v$,
moves at least five vertices from $R$ into $B$, each losing at least $\Delta$, so
\begin{equation}\label{eq:mhigh}
C(\mu)\le C(\mu-\gamma_{\mathrm w}-5\Delta)+C(\mu-\gamma_{\mathrm w}).
\end{equation}

Under the analysis weights in Table~\ref{table-weight2}, none of the above recurrences
has a branching factor greater than 1.9985.
Since initially $\mu\le n_0=n-|T|$, we get

\begin{theorem}\label{thm:weighted}
\textsc{Vertex-Weighted Directed Multiway Cut} can be solved deterministically in
$O^*(1.9985^{n-|T|})$ time and polynomial space.
\end{theorem}

\section{Conclusion and Discussion}
Fomin et al.~\cite{DBLP:journals/jacm/FominGLS19} put forth a general method for a range of deletion problems to design exact algorithms and surpass the $2^n$ barrier by leveraging single-exponential parameterized algorithms. Given an $O^*(a^k)$-time parameterized algorithm where $k$ is the solution size, we can derive an $O^*(b^n)$-time algorithm where $b=2-1/a$. For \textsc{Directed Multiway Cut}, we have not found a single-exponential parameterized algorithm yet; for \textsc{Directed Multicut}, the problem even becomes W[1]-hard when there are at least four terminal pairs~\cite{DBLP:conf/soda/PilipczukW16}. The recent $O^*(k^{O(k)})$-time parameterized algorithm of Chu et al.~\cite{ChuEtAl2026} for \textsc{Directed Multiway Cut} does not give a single-exponential $O^*(a^k)$ bound for a constant $a$. Consequently, the method in~\cite{DBLP:journals/jacm/FominGLS19} cannot be currently applied to these two problems.
Fortunately, we have managed to break $2^n$ for \textsc{Directed Multiway Cut} by proposing a relatively simple algorithm.
It seems that directed problems are much harder than undirected problems. One of the most important techniques in this paper to achieve the breakthrough should be the technique of reversing the direction of edges. Before reversing the direction, we use some branching rules; after reversing the direction, we use some other branching rules.
This technique, not being used before as far as we know, makes the algorithm simple.
The algorithm presented in this paper may aid us in understanding exhaustive-searching algorithms for related cut problems in directed graphs.

\section*{Acknowledgements}
An initial version of this paper was reported at the 2024 European Symposium on Algorithms (ESA 2024)~\cite{xiaoESA}. Grammarly was used in this paper for grammar checking and revision of some sentences. Recently, GPT-5.6 Sol was used to check the correctness of the paper once again.


\bibliographystyle{plain}
\bibliography{mwc}

@article{mwc,
  author       = {Elias Dahlhaus and
                  David S. Johnson and
                  Christos H. Papadimitriou and
                  Paul D. Seymour and
                  Mihalis Yannakakis},
  title        = {The Complexity of Multiterminal Cuts},
  journal      = {{SIAM} J. Comput.},
  volume       = {23},
  number       = {4},
  pages        = {864--894},
  year         = {1994}
}

@article{d-mwc,
  author       = {Naveen Garg and
                  Vijay V. Vazirani and
                  Mihalis Yannakakis},
  title        = {Multiway cuts in node weighted graphs},
  journal      = {J. Algorithms},
  volume       = {50},
  number       = {1},
  pages        = {49--61},
  year         = {2004}
}

@article{multicut1,
  author       = {Gruia C{\u{a}}linescu and
                  Cristina G. Fernandes and
                  Bruce A. Reed},
  title        = {Multicuts in unweighted graphs and digraphs with bounded degree and
                  bounded tree-width},
  journal      = {J. Algorithms},
  volume       = {48},
  number       = {2},
  pages        = {333--359},
  year         = {2003}
}

@inproceedings{multicutESA,
  author       = {Daniel Lokshtanov and
                  Saket Saurabh and
                  Ondrej Such{\'{y}}},
  editor       = {Andreas S. Schulz and
                  Dorothea Wagner},
  title        = {Solving Multicut Faster Than $2^n$},
  booktitle    = {Algorithms - {ESA} 2014 - 22nd Annual European Symposium, Wroclaw,
                  Poland, September 8-10, 2014. Proceedings},
  series       = {Lecture Notes in Computer Science},
  volume       = {8737},
  pages        = {666--676},
  publisher    = {Springer},
  year         = {2014}
}

@article{Xiao10,
  author       = {Mingyu Xiao},
  title        = {Simple and Improved Parameterized Algorithms for Multiterminal Cuts},
  journal      = {Theory Comput. Syst.},
  volume       = {46},
  number       = {4},
  pages        = {723--736},
  year         = {2010}
}

@article{CyganPPW13,
  author       = {Marek Cygan and
                  Marcin Pilipczuk and
                  Michal Pilipczuk and
                  Jakub Onufry Wojtaszczyk},
  title        = {On multiway cut parameterized above lower bounds},
  journal      = {{ACM} Trans. Comput. Theory},
  volume       = {5},
  number       = {1},
  pages        = {3:1--3:11},
  year         = {2013}
}

@article{ChenLL09,
  author       = {Jianer Chen and
                  Yang Liu and
                  Songjian Lu},
  title        = {An Improved Parameterized Algorithm for the Minimum Node Multiway
                  Cut Problem},
  journal      = {Algorithmica},
  volume       = {55},
  number       = {1},
  pages        = {1--13},
  year         = {2009}
}

@article{DBLP:journals/jacm/FominGLS19,
  author       = {Fedor V. Fomin and
                  Serge Gaspers and
                  Daniel Lokshtanov and
                  Saket Saurabh},
  title        = {Exact Algorithms via Monotone Local Search},
  journal      = {J. {ACM}},
  volume       = {66},
  number       = {2},
  pages        = {8:1--8:23},
  year         = {2019}
}

@article{terminalP,
  author       = {Rajesh Chitnis and
                  Fedor V. Fomin and
                  Daniel Lokshtanov and
                  Pranabendu Misra and
                  M. S. Ramanujan and
                  Saket Saurabh},
  title        = {Faster exact algorithms for some terminal set problems},
  journal      = {J. Comput. Syst. Sci.},
  volume       = {88},
  pages        = {195--207},
  year         = {2017}
}

@article{DBLP:journals/algorithmica/FominHKPV14,
  author       = {Fedor V. Fomin and
                  Pinar Heggernes and
                  Dieter Kratsch and
                  Charis Papadopoulos and
                  Yngve Villanger},
  title        = {Enumerating Minimal Subset Feedback Vertex Sets},
  journal      = {Algorithmica},
  volume       = {69},
  number       = {1},
  pages        = {216--231},
  year         = {2014}
}

@book{DBLP:series/txtcs/FominK10,
  author       = {Fedor V. Fomin and
                  Dieter Kratsch},
  title        = {Exact Exponential Algorithms},
  series       = {Texts in Theoretical Computer Science. An {EATCS} Series},
  publisher    = {Springer},
  year         = {2010}
}

@article{DBLP:conf/soda/PilipczukW16,
  author       = {Marcin Pilipczuk and
                  Magnus Wahlstr{\"{o}}m},
  title        = {Directed Multicut is W[1]-hard, Even for Four Terminal Pairs},
  journal      = {{ACM} Trans. Comput. Theory},
  volume       = {10},
  number       = {3},
  pages        = {13:1--13:18},
  year         = {2018},
  url          = {https://doi.org/10.1145/3201775},
  doi          = {10.1145/3201775},
  bibsource    = {dblp computer science bibliography, https://dblp.org}
}

@article{eQP,
  author       = {David Eppstein},
  title        = {Quasiconvex analysis of multivariate recurrence equations for backtracking
                  algorithms},
  journal      = {{ACM} Trans. Algorithms},
  volume       = {2},
  number       = {4},
  pages        = {492--509},
  year         = {2006},
  url          = {https://doi.org/10.1145/1198513.1198515},
  doi          = {10.1145/1198513.1198515},
  bibsource    = {dblp computer science bibliography, https://dblp.org}
}

@article{DBLP:journals/jacm/FominGK09,
  author       = {Fedor V. Fomin and
                  Fabrizio Grandoni and
                  Dieter Kratsch},
  title        = {A measure {\&} conquer approach for the analysis of exact algorithms},
  journal      = {J. {ACM}},
  volume       = {56},
  number       = {5},
  pages        = {25:1--25:32},
  year         = {2009},
  url          = {https://doi.org/10.1145/1552285.1552286},
  doi          = {10.1145/1552285.1552286},
  bibsource    = {dblp computer science bibliography, https://dblp.org}
}

@article{DBLP:journals/siamcomp/NaorZ01,
  author       = {Joseph Naor and
                  Leonid Zosin},
  title        = {A 2-Approximation Algorithm for the Directed Multiway Cut Problem},
  journal      = {{SIAM} J. Comput.},
  volume       = {31},
  number       = {2},
  pages        = {477--482},
  year         = {2001},
  url          = {https://doi.org/10.1137/S009753979732147X},
  doi          = {10.1137/S009753979732147X},
  bibsource    = {dblp computer science bibliography, https://dblp.org}
}

@inproceedings{DBLP:conf/soda/EsmerKMNS24,
  author       = {Baris Can Esmer and
                  Ariel Kulik and
                  D{\'{a}}niel Marx and
                  Daniel Neuen and
                  Roohani Sharma},
  editor       = {David P. Woodruff},
  title        = {Optimally Repurposing Existing Algorithms to Obtain Exponential-Time
                  Approximations},
  booktitle    = {Proceedings of the 2024 {ACM-SIAM} Symposium on Discrete Algorithms,
                  {SODA} 2024, Alexandria, VA, USA, January 7-10, 2024},
  pages        = {314--345},
  publisher    = {{SIAM}},
  year         = {2024},
  url          = {https://doi.org/10.1137/1.9781611977912.13},
  doi          = {10.1137/1.9781611977912.13},
  bibsource    = {dblp computer science bibliography, https://dblp.org}
}

@inproceedings{xiaoESA,
  author       = {Mingyu Xiao},
  editor       = {Timothy M. Chan and
                  Johannes Fischer and
                  John Iacono and
                  Grzegorz Herman},
  title        = {Solving Directed Multiway Cut Faster Than $2^n$},
  booktitle    = {32nd Annual European Symposium on Algorithms, {ESA} 2024, Royal Holloway,
                  London, United Kingdom, September 2-4, 2024},
  series       = {LIPIcs},
  volume       = {308},
  pages        = {104:1--104:13},
  publisher    = {Schloss Dagstuhl - Leibniz-Zentrum f{\"{u}}r Informatik},
  year         = {2024},
  url          = {https://doi.org/10.4230/LIPIcs.ESA.2024.104},
  doi          = {10.4230/LIPICS.ESA.2024.104},
  bibsource    = {dblp computer science bibliography, https://dblp.org}
}

@inproceedings{ChuEtAl2026,
  author = {Huairui Chu and Yuxi Liu and Daniel Lokshtanov and Junqiang Peng and Kangyi Tian and Mingyu Xiao},
  title = {{Faster Parameterized Vertex Multicut}},
  booktitle = {Proceedings of the 2027 {ACM-SIAM} Symposium on Discrete Algorithms, {SODA} 2027},
  publisher = {{SIAM}},
  year = {2027},
  note = {To appear. Preprint: \url{https://arxiv.org/abs/2602.13981}}
}

\end{document}